\documentclass[12pt]{iopart}

\usepackage{iopams}
\usepackage{graphicx}
\begin{document}

\title[]{Implementation of adiabatic Abelian geometric gates with
superconducting phase qubits}

\author{Z.~H.~Peng$^1$, H.~F.~Chu$^1$, Z.~D.~Wang$^2$, and D.~N.~Zheng$^1$}

\address{$^1$ National Laboratory for Superconductivity, Institute of
Physics and Beijing National Laboratory for Condensed Matter
Physics,Chinese Academy of Sciences, Beijing 100080, PR China}

\address{$^2$ Department of Physics and
Center of Theoretical and Computational Physics, The University of
Hong Kong, Pokfulam Road, Hong Kong, PR China}
\ead{zwang@hkucc.hku.hk}
\ead{dzheng@ssc.iphy.ac.cn}

\date{\today }

\begin{abstract}
We have developed an adiabatic Abelian geometric quantum computation
strategy based on the non-degenerate energy eigenstates in (but not
limited to) superconducting phase-qubit systems. The fidelity of the
designed quantum gate was evaluated in the presence of simulated
thermal fluctuation in superconducting phase qubits circuit and was
found to be rather robust against the random errors. In addition, it
was elucidated that the Berry phase in the designed adiabatic
evolution may be detected directly via the quantum state tomography
developed for superconducting qubits.

\end{abstract}

\maketitle

\tableofcontents

\section{Introduction}
Quantum computation (QC) has been attracting more and more interests
for the past decade due to its unrivaled power exceeds that of the
classical counterpart in solving certain problems. Significant and
exciting progress has been achieved both theoretically and
experimentally in this field. Nevertheless, there are still many
difficulties and challenges in physical implementation of quantum
computation. To increase the fidelity of quantum gates to an
acceptable high level is one of them, and is essential to construct
workable quantum logical gates in scalable quantum computers.
Recently, several promising schemes based on the geometric phase
have been proposed for achieving built-in fault-tolerant quantum
gates with high
fidelities\cite{Berry,Falci,Peng,Jones,hqc,duan,zhangp,ZW,Zhang,Zanardi,Zhu2006},
as it is believed that QC based on geometric phase shifts may be
more robust against certain type of stochastic errors in the
control/operation parameters/processes than dynamic quantum gates.
For the past several years, based on the adiabatic non-Abelian
geometric phase shifts accumulated for degenerate/dark energy
eigenstates, the
holonomic quantum computation (HQC) has been proposed and developed~\cite%
{hqc,duan,zhangp}. In addition, a geometric QC scheme based
on the nonadiabatic but cyclic geometric phases has been achieved~\cite%
{ZW}. On the other hand, the environment effects on the Berry
phase of a two-level system have also been
addressed\cite{DeChiara}.

On the other hand, it has been realized that superconducting qubits
 provide us a
promising approach towards a scalable solid-state quantum computer
\cite{Makhlin,Nakamura,Chiorescu,Yu,Martinis02}. However, the
unavoidable random errors may lead to a serious reduction of the
fidelities of the wanted quantum gates. Constructing fault-tolerant
quantum logic gates in superconducting based on geometric phase has
been paid particular attention recently \cite{Falci,Peng,zhangp,ZW}.
Being different from the above schemes, we here develop an adiabatic
Abelian geometric QC scheme based on the non-degenerate energy
eigenstates (considering qubit-qubit interaction as
\textbf{$\sigma_{y}\sigma_{y}$} type), and then analyze its fidelity
against certain kind of simulated noises in superconducting phase
qubits.

The paper is organized as follows. In section
\ref{Section:AdiabaticQG}, we introduce the adiabatic Abelian
geometric gates. In section \ref{section:fidelity}, we simulate the
fidelities of both the single-qubit and the two-qubit controlled
adiabatic Abelian geometric gate in the presence of random
fluctuations. In section \ref{section:detection}, the detection of
the Berry phase in superconducting phase qubits is analyzed. Section
\ref{section:discussion} presents relevant discussions and a brief
summary.
\section{Adiabatic Abelian Geometric Gates}

\label{Section:AdiabaticQG} We now elaborate our adiabatic Abelian
geometric QC strategy. A qubit system, when the system Hamiltonian
with two instantaneous non-degenerate energy levels changes
adiabatically and cyclically in a parameter space with the period
$\tau $, behaves like a spin$\frac{1}{2}$ particle in a magnetic
field, with the Hamiltonian as $H= {\vec \sigma} \cdot
\mathbf{B}/2$. Under the adiabatic approximation, the two orthogonal
energy eigenstates $|\psi _{\pm }(t)\rangle $ will also follow the
Hamiltonian to evolve adiabatically and cyclically starting from the
initial states $|\psi
_{\pm }(0)\rangle $: 
$|\psi _{\pm }(\tau )\rangle =U(\tau )|\psi _{\pm }(0)\rangle
\approx \exp (\pm i\gamma )|\psi _{\pm }(0)\rangle $, where the
$U(\tau )$ is the evolution operator of the system and the $\pm
\gamma $ are respectively the total phases accumulated for the
$|\psi _{\pm }\rangle $ states in the evolution. If we denote $|\psi
_{+}(0)\rangle =\cos \frac{\xi }{2}|0\rangle +e^{i\eta }\sin
\frac{\xi }{2}|1\rangle $ and $|\psi _{-}(0)\rangle =-\sin
\frac{\xi }{2}|0\rangle +e^{i\eta }\cos \frac{\xi }{2}|1\rangle $, with $%
|0\rangle $ and $|1\rangle $ as the two eigenstates of $\sigma
_{z}$ and being chosen as our computational basis ($\eta =0$ for
$B_{y}(0)=0$ and $\eta =\pi/2$ for $B_{x}(0)=0$).

Thus, for an arbitrary initial state of the system $|\psi
_{in}\rangle =a_{+}|\psi _{+}(0)\rangle +a_{-}|\psi _{-}(0)\rangle $
with $a_{\pm }=\langle \psi _{\pm }(0)|\psi _{in}\rangle $, after
the adiabatic and cyclic evolution time $\tau $, the final state is
found to be $|\psi _{f}\rangle \approx U(\gamma, \xi, \eta
)|\psi _{in}\rangle $, where 
\begin{equation}
\label{equation:single} U=\left(
\begin{array}{cc}
e^{i\gamma }\cos ^{2}\frac{\xi }{2}+e^{-i\gamma }\sin ^{2}\frac{\xi
}{2} &
ie^{-i\eta }\sin \xi \sin \gamma  \\
ie^{i\eta }\sin \xi \sin \gamma  & e^{i\gamma }\sin ^{2}\frac{\xi }{2}%
+e^{-i\gamma }\cos ^{2}\frac{\xi }{2}%
\end{array}%
\right) .
\end{equation}
Moreover, a controlled two-qubit gate may also be achieved under
the condition that the control qubit is off resonance in the
operation of the target qubit (to be addressed later).

 Considering that $\gamma$ is the
total phase usually consisting of both geometric and dynamic phases,
we here illustrate how to eliminate the corresponding dynamic
phase in a simple two-loop quantum gate operation, so that the achieved $U$%
-gate is a pure geometric one depending only on the geometric phase
accumulated in the whole evolution. Set the basic adiabatically
cyclic evolution time to be $\tau_0$ with the corresponding
geometric Berry phase
as $\gamma^0_g$. After the first cyclic evolution of the states $%
|\psi_{\pm}\rangle$ by driving the fictitious field adiabatically
with the period $\tau_0$, we reverse promptly the fictitious field
direction such
that the states $|\psi_{\pm}\rangle$ are unchanged, i.e., $\mathbf{B}%
(\tau_0+0)=-\mathbf{B}(\tau_0)$ and $|\psi_{\pm}(\tau_0+0)
\rangle=|\psi_{\pm}(\tau_0)\rangle$. Then we let $\mathbf{B}(\tau_0+t)=-%
\mathbf{B}(t)$ in the second $\tau_0$-time cycle evolution. During
the second period, the state $|\psi_{+}\rangle$ ($|\psi_{-}\rangle$)
acquires the same geometric phase as that in the first period but
with the reversal
sign of the dynamic phase, so that the accumulated total phase of $%
|\psi_{+}\rangle$ ($|\psi_{-}\rangle$) at the end second period is a
pure geometric phase $\gamma=2\gamma^0_g$ ($-2\gamma^0_g$).
Therefore, the pure geometric quantum $U(2\tau_0)$-gates given by
equation (\ref{equation:solid}) can be obtained. For example, two
simple noncommutable single-qubit gates, the type of
Hadamard-gate and the type of NOT gate, can be achieved by setting ($%
\xi=\pi/4$, $\eta=0$, $\gamma^0_g=\pi/4$) and  ($\xi=\pi/2$, $\eta=0$, $%
\gamma^0_g=\pi/4$), respectively.

\section{Fidelity of Adiabatic Abelian gates}
\label{section:fidelity}
 Recently, it was reported from numerical simulations that the
earlier proposed two kinds of geometric quantum gates, a class of
non-Abenian
holonomic gates~\cite{hqc} and a set of nonadiabatic Abelian geometric gates~%
\cite{ZW,Zhang}, are likely more robust against stochastic control
errors than dynamical gates~\cite{Zanardi,Zhu2006}. It is natural to
ask whether the present adiabatic Abelian geometric gates are also
robust against stochastic errors as expected. To answer this
question, here we illustrate this by superconducting phase qubits.

\begin{figure}
\centering
\includegraphics[width=0.4\textwidth]{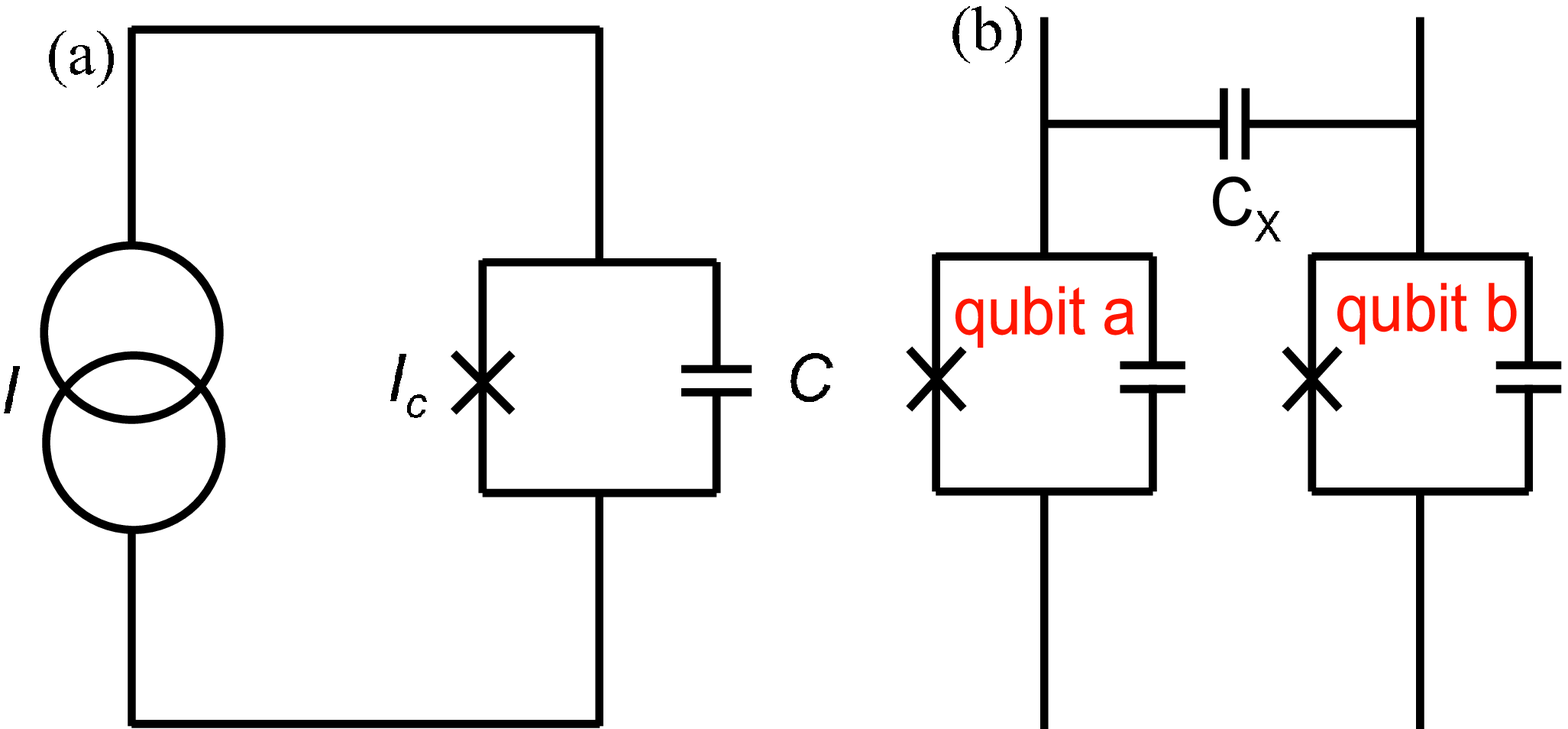}

\includegraphics[width=0.4\textwidth]{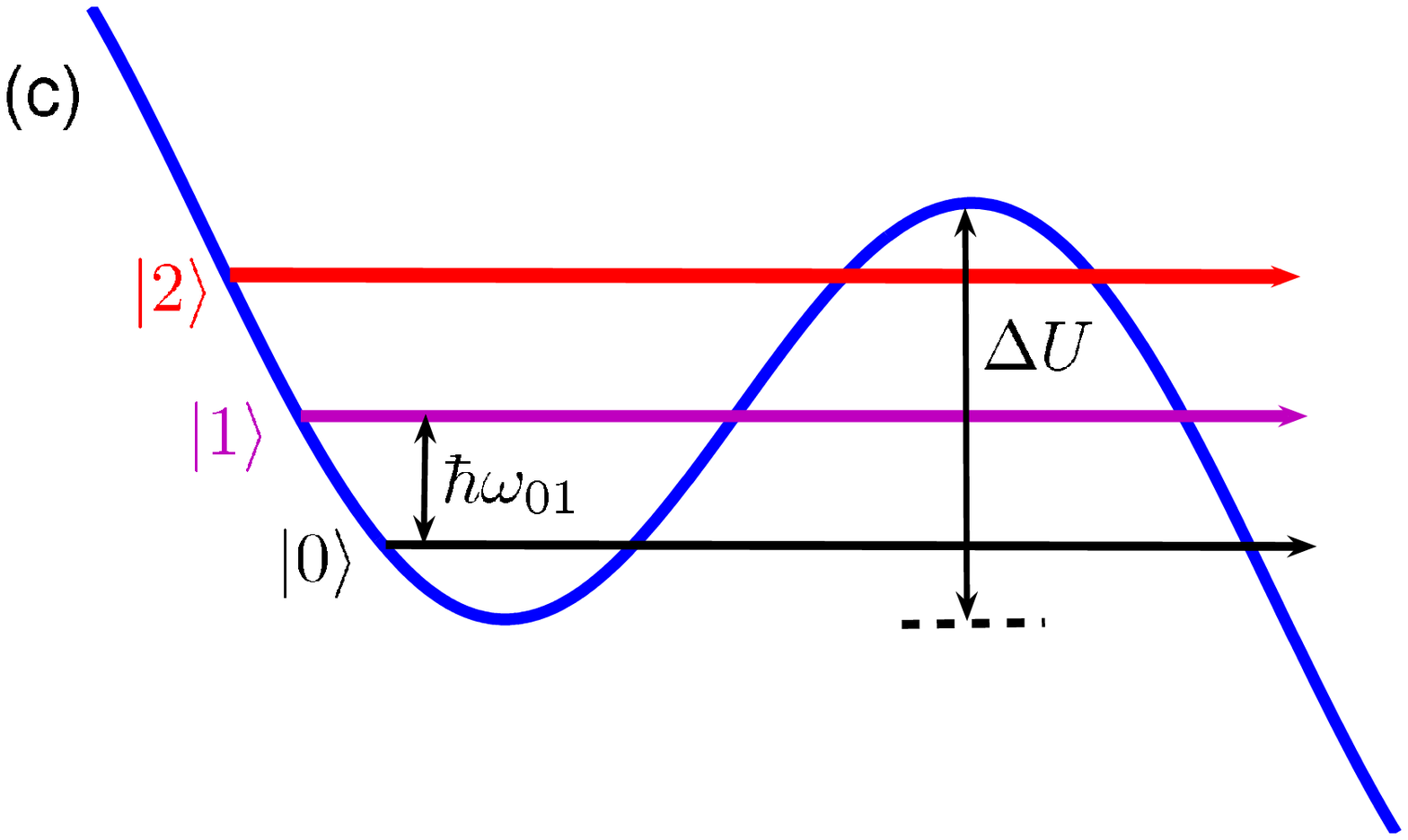}
\caption{{\protect\footnotesize} Schematic diagrams of (a) a circuit
of phase qubit; (b) a two-qubit gate, where two single phase qubits
are coupled by a capacitor $C_{x}$; (c) quantized energy levels in a
current biased Josephson Junction, where the two lowest eigenstates
$|0\rangle$ and $|1\rangle$ form a qubit.} \label{fig:circuit}
\end{figure}

As is known,  a large current-biased Josephson
junction(figure~\ref{fig:circuit}a,~c) may work as a typical phase
qubit, which can be considered as an anharmonic $LC$ resonator with
resonance frequency $\omega_{p}=(L_{J}C_{J})^{-1/2}$, whose two
lowest quantized engergy levels are chosen as the qubit states\cite%
{Yu,Martinis02,Clarke}, where $L_{J}$ is the Josephson inductance
and $C_{J}$ is the junction capacitance.
The Josephson inductance is given by $%
L_{J}=\phi_{0}/2\pi{I_{c}\cos\delta}$, where $I_{c}$ is the junction
critical current, $\delta$ is the phase difference across the
junction given through $I=I_{c}\sin\delta$, and $\phi_{0}=h/2e$ is
the superconducting flux
quantum. As the junction bias current $I$ close to the critical current $%
I_{c}$, the anharmonic potential may be approximated by a cubic
potential
parameterized by the potential barrier height $\Delta U(I)=(2\sqrt{2}%
I_{c}\phi_{0}/3\pi)[1-I/I_{c}]^{3/2}$ and a plasma oscillation
frequency at the bottom of the well $\omega_{p}(I)=2^{1/4}(2\pi
I_{c}/\phi_{0}C)^{1/2}[1-I/I_{0}]^{1/4}$. Microwave induces
transitions
between levels at a frequency $\omega_{mn}=E_{mn}/\hbar=(E_{m}-E_{n})/\hbar$%
, where $E_{n}$ is the energy of state $|n\rangle$. The state of the
qubit
can be controlled with dc and microwave pulses of bias current $%
I(t)=I_{dc}+\delta I_{dc}(t)+I_{\mu w}(t)\cos\phi
\cos\omega_{10}t+I_{\mu w}(t)\sin\phi\sin\omega_{10}t$. As usual,
under a reasonable approximation that the dynamics of the system is
restricted to the Hilbert space spanned by the lowest two states,
the Hamiltonian in the $\omega_{10}$ rotating frame may be written
as
\begin{eqnarray}
\label{equation:hamiltonian} H=\hat{\sigma}_{x}I_{\mu
w}(t)\cos\phi\sqrt{\hbar/2\omega_{10}C}/2+\hat{\sigma}_{y}I_{\mu
w}(t)\sin\phi\sqrt{\hbar/2\omega_{10}C}/2 \nonumber
\\ ~~~~~+\hat{\sigma}_{z}\delta I_{dc}(t)(\partial E_{10}/\partial
I_{dc})/2,
\end{eqnarray}
where $\hat{\sigma}_{x,y,z}$ are Pauli operators. As schematically
shown in figure~\ref{fig:circuit}b, a untrivial two-qubit gate could
be constructed by capacitive coupling.

From equation (\ref{equation:hamiltonian}), one could define a fictitious field $\mathbf{B}\equiv(\nu\cos\phi,\nu%
\sin \phi,\Delta\omega)$, where $\nu=I_{\mu w}(t)\sqrt{\hbar/2\omega_{10}C}%
,\Delta\omega=\delta I_{dc}(t)(\partial E_{10}/\partial I_{dc})$.
The phase qubit thus behaves like a spin-$\frac{1}{2}$ particle in a
magnetic field, with the Hamiltonian as $H= {\vec \sigma} \cdot
\mathbf{B}/2$. For such a quantum system, the acquired geometric
phase of its energy eigenstate is equal to half of the solid angle
subtended by the area in the parameter space enclosed by the closed
evolution loop of the fictitious magnetic field. The solid angle may
be evaluated by~\cite{ZW}
\begin{equation}
\label{equation:solid} \Omega=\int_{0}^\tau
\frac{B_{x}\partial_{t}B_{y}-B_{y}\partial_{t}B_{x}}{|B|(B_{z}+|B|)}dt
,
\end{equation}
under the condition $\mathbf{B}(\tau)=\mathbf{B}(0)$. Especially,
when the adiabatic evolution path forms a cone in the parameter
space $\{\mathbf{B}\}$ under the varying current, the corresponding
Berry phases of two energy eigenstates are simply given
by~\cite{Berry}
$\gamma_g=\pm
\pi[1-\Delta\omega/\sqrt{(\Delta\omega)^{2}+(\nu)^{2}}]$.

We will perform ceratin kind of numerical simulations on the
fidelity of the adiabatic Berry phase gates given by
equation~(\ref{equation:single}) and (\ref{equation:control}),
subject to the modelled random noises for the weakly fluctuated
driving bias current. Note that, in the numerical studies of
Refs.\cite{Zanardi,Zhu2006}, the fluctuations of control parameters
were assumed to be uniformly distributed in an interval and merely a
certain type of states in the Bloch sphere were sampled to evaluate
the average fidelity of gates. While in real experiments, the finite
impedance of the bias-current source produces the decoherence of
superconducting phase qubit from the dissipation and noises. We here
mainly consider the noise in the current due to the thermal
fluctuation, which is likely one of the main noise sources in
superconducting qubits circuit\cite{Bertet}. The actual noise
current generated by a resistance $R$ at temperature T may be
estimated by $I_{n}(rms)=(4k_B TB/R)^{1/2}$, where  $B$ is the
bandwidth parameter~\cite{Horowitz}. The amplitude of the noise
current would obey a Gaussian distribution\cite{Martinis02}. In
fact, assuming the critical current of superconducting phase qubits
$I_{c}\sim10\mu $A\cite{Steffen}, the measurement bandwidth
$B\sim10$GHz and the bias resistor R$\sim$10K$\Omega$ at
T$\sim$4.2K, the total current noise would be around 15 nA. The bias
current is driven close to the critical current $I_{c}$ and the
transition frequency between qubit states is
$\omega_{10}/2\pi\sim6$GHz. The Rabi frequency is
$\nu/2\pi\sim$300MHz and the Ramsey frequency is
$\Delta\omega/2\pi\sim300$MHz. The fluctuation of Ramsey frequency
resulting from noise in the bias current is about 10MHz.
We will below evaluate  the average fidelity of the designed new
geometric quantum gates subject to this type of errors for any input
state.

As is known, the average fidelity of a quantum logic gate in the
presence of random noises may be defined as
\begin{equation}
\overline{F}= \lim_{N\rightarrow\infty}
\frac{1}{N}\sum^{N}_{j=1}
|\langle\psi_{in}|\widehat{U}^{\dag}\widehat{U}^{j}_{noise}| \psi
_{in}\rangle|^{2}
,
\end{equation}
where $|\psi_{in}\rangle=[\cos(\theta_i/2), e^{i\varphi_i}
\sin(\theta_i/2)]^{T}$($T$ represents the transposition of matrix.), $\theta_i\in$ [0,$\pi$] and $\varphi_i\in$ [0,2$\pi$%
] are the coordinators of the input state in our numerical simulations. Here, $%
U$ is an ideal adiabatic quantum gate denoted by equation (\ref{equation:solid}) in the absence of random errors and $%
U_{noise}$ is the gate operator in the presence of random errors.

For simplicity, we here focus only on a cone-type adiabatic evolution: $%
\xi=\tan^{-1}(\nu/\Delta \omega)$. Since this type of adiabatic
evolution of the field could have various forms, it seems subtle to
directly simulate the state evolution with a reliable way under the
adiabatic condition in the presence of random errors. To evade this
subtle issue, we adopt a simple method to model effectively the
effect of random errors occurred in the evolution. For a given
configuration of errors in the evolution, let us look
at the final state $|\tilde{\psi}_{+}\rangle=U_{noise}|\psi_{+}%
\rangle(0)$ and regard it to be evolved adiabatically and
cyclically as well as ideally from a visual initial state $|\tilde{\psi}_{+}\rangle(0)= \cos%
\frac{\tilde{\xi}}{2}|0\rangle+e^{i\tilde{\eta}}\sin\frac{\tilde{\xi}}{2}%
|1\rangle=U^{-1}(\tilde{\gamma}, \tilde{\xi},
\tilde{\eta})|\tilde{\psi}_{+}\rangle$,
namely, $|\tilde{\psi}_{+}\rangle=e^{i\tilde{\gamma}}|%
\tilde{\psi}_{+}\rangle(0)$. In this sense, $U_{noise}$ can be expressed as $%
U(\tilde{\gamma}, \tilde{\xi}, \tilde{\eta})$ in equation
(\ref{equation:solid}) with $\tilde{\gamma}$ as the geometric Berry
phase of the two loops. Here, the random parameters
$(\tilde{\gamma}, \tilde{\xi})$ may be determined by the
randomly fluctuated bias current from the relations $\gamma^0_g$ ($\nu$, $%
\Delta \omega$) and $\xi$ ($\nu$, $\Delta \omega$), with the
Gaussian-type error probability density
\begin{equation}
\frac{dp(x)}{dx}=\frac{\exp(-x^{2}/2\sigma^{2})}{\sqrt{2\pi}\sigma},
\end{equation}
where the $x$ is the deviation from $\nu$ (or $\Delta\omega$,
$\eta$), and $\sigma$ is the mean squared noise.
\begin{figure}[tbp]
\centering
\includegraphics[width=0.5\textwidth]{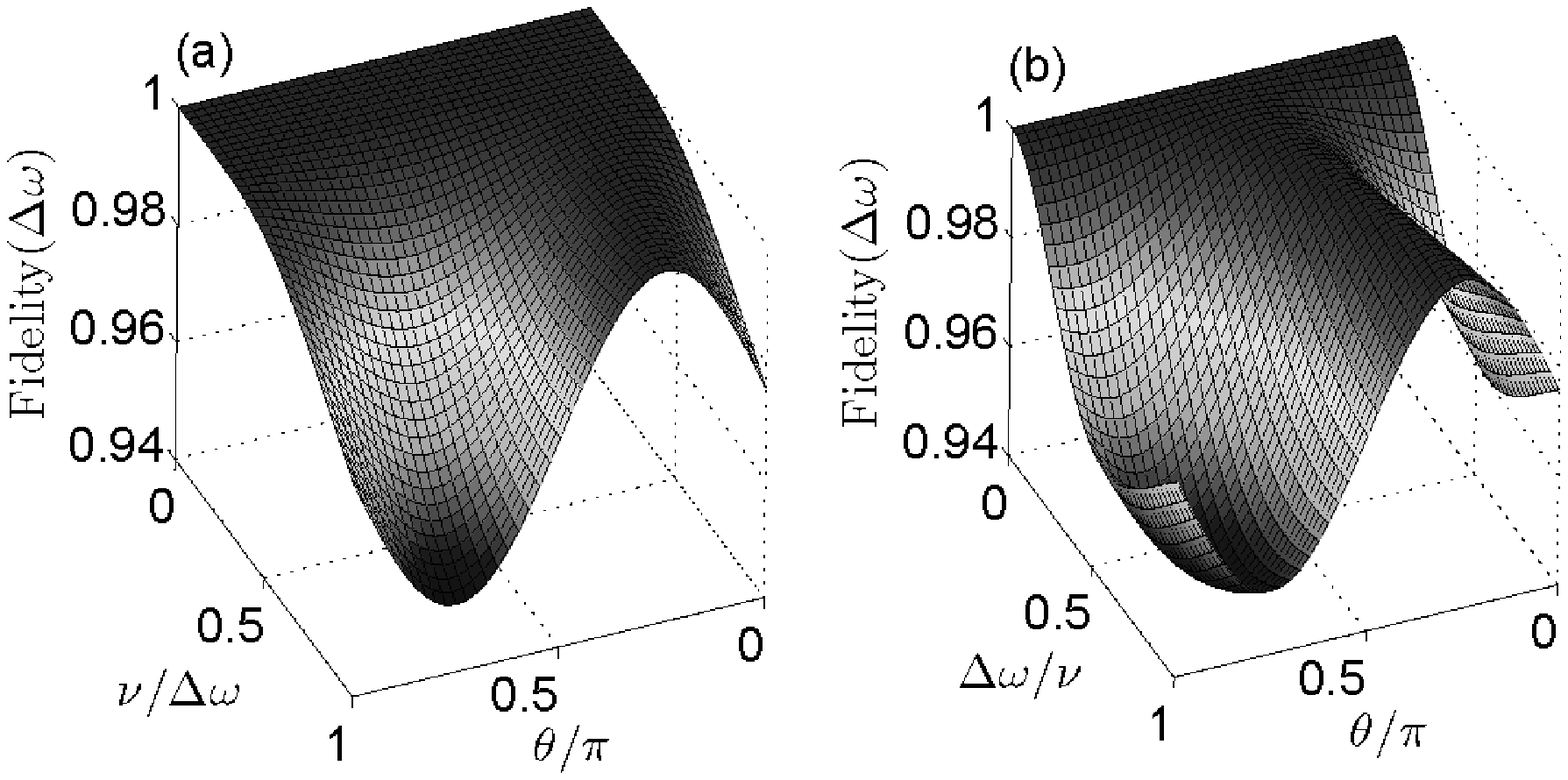} %

\includegraphics[width=0.5\textwidth]{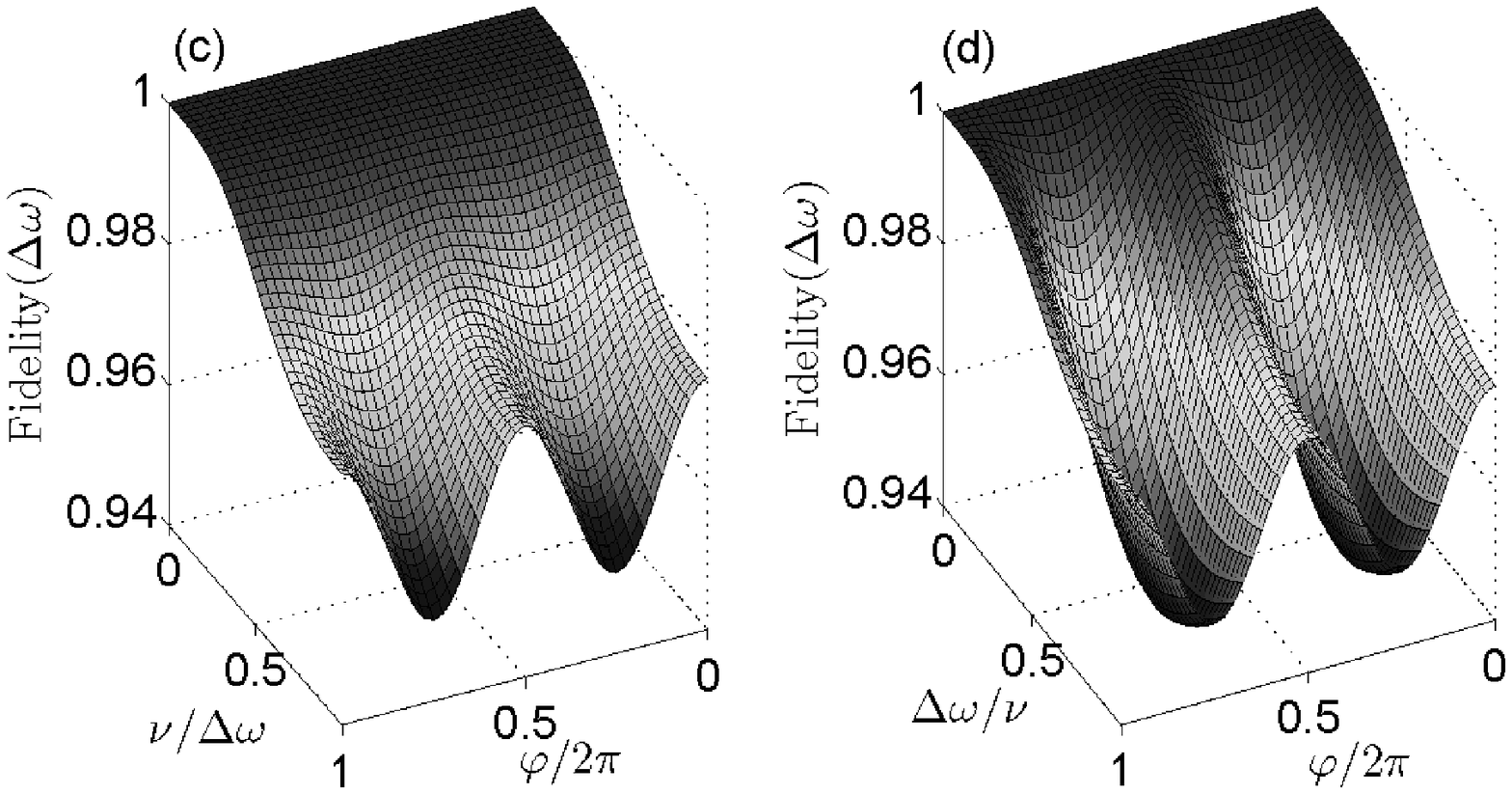}
\caption{{\protect\footnotesize The fidelity of single-qubit gate in
 the presence of
 $\Delta\protect\omega$-fluctuations, where
  $\protect\varphi_{i}=0$ in (a) and (b), $\protect\theta_{i}=\protect%
\pi/2$ in (c) and (d). Parameters are: $\omega_{10}/2\pi=6$GHz and
$\protect\sigma_{0}=0.1$.}} \label{fig:delta_omega}
\end{figure}

\begin{figure}[tbp]
\centering
\includegraphics[width=0.5\textwidth]{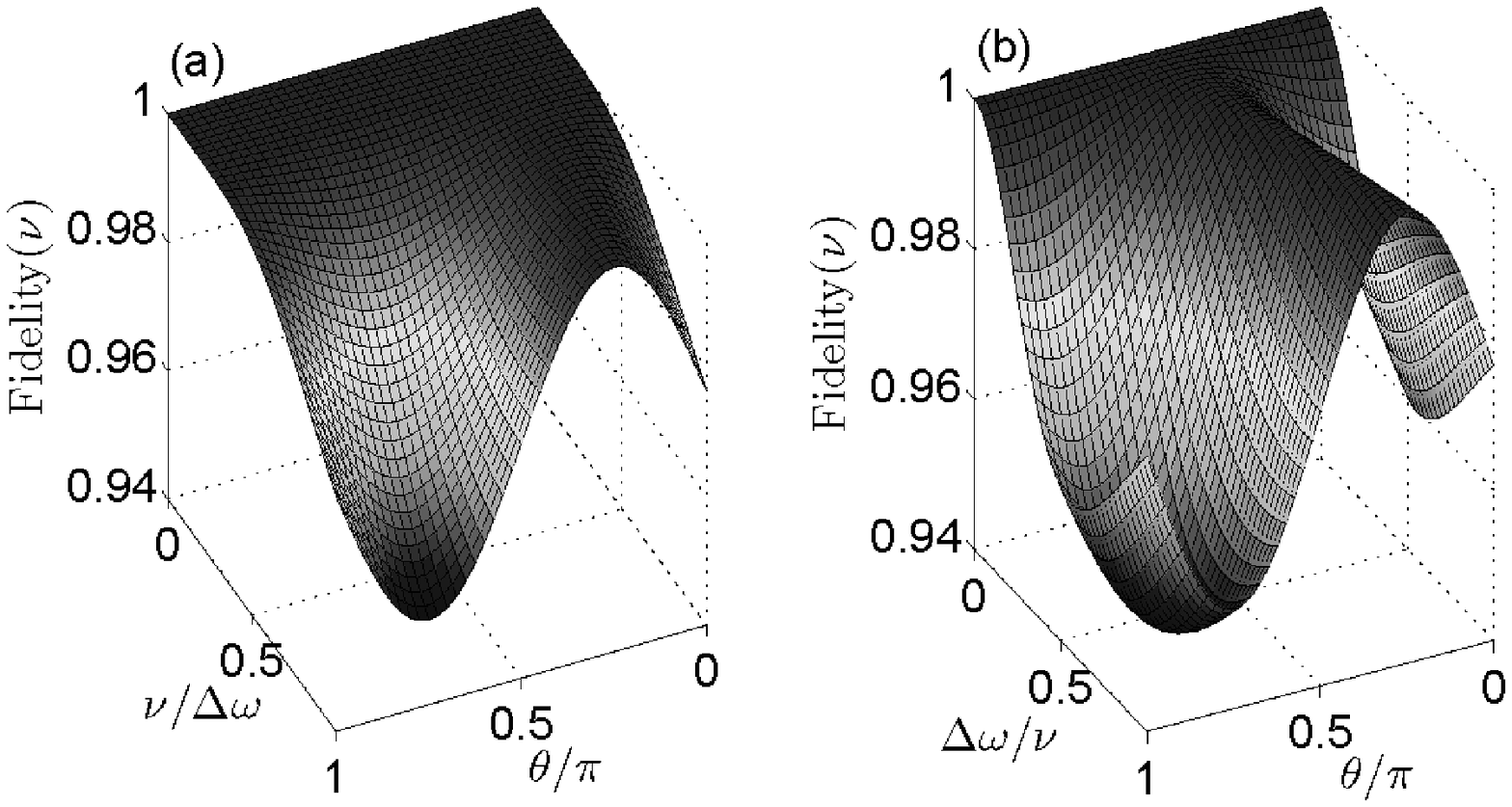} %

\includegraphics[width=0.5\textwidth]{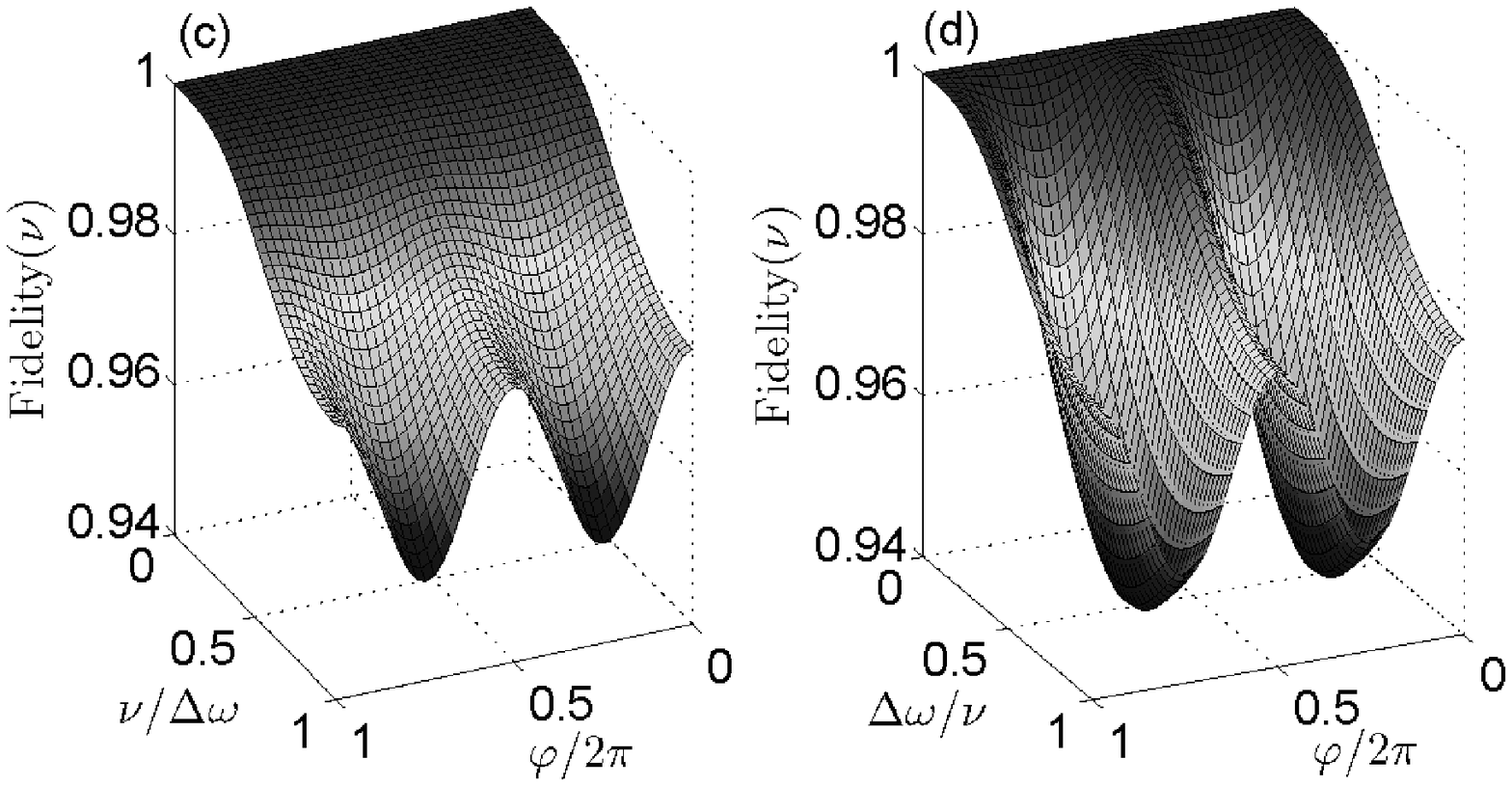}
\caption{{\protect\footnotesize The fidelity of single-qubit gate in
 the presence of
$\protect\nu$-fluctuations, where $%
\protect\varphi_{i}=0$ in (a) and (b),
$\protect\theta_{i}=\protect\pi/2$ in (c) and (d). Parameters are:
$\omega_{10}/2\pi=6$GHz and $\protect\sigma_{1}=0.1$.}}
\label{fig:nu}
\end{figure}

\begin{figure}[tbp]
\centering
\includegraphics[width=0.5\textwidth]{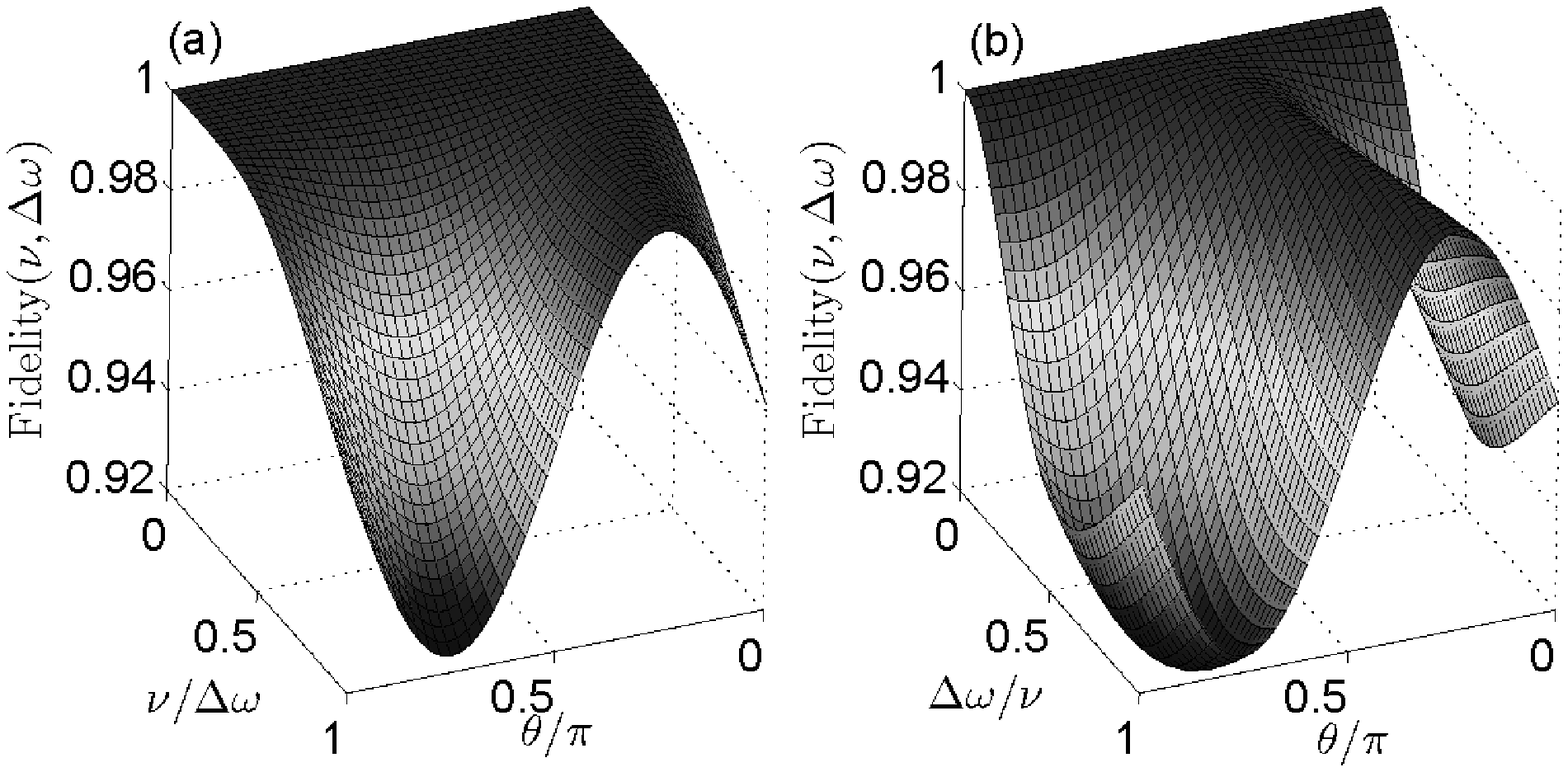} %

\includegraphics[width=0.5\textwidth]{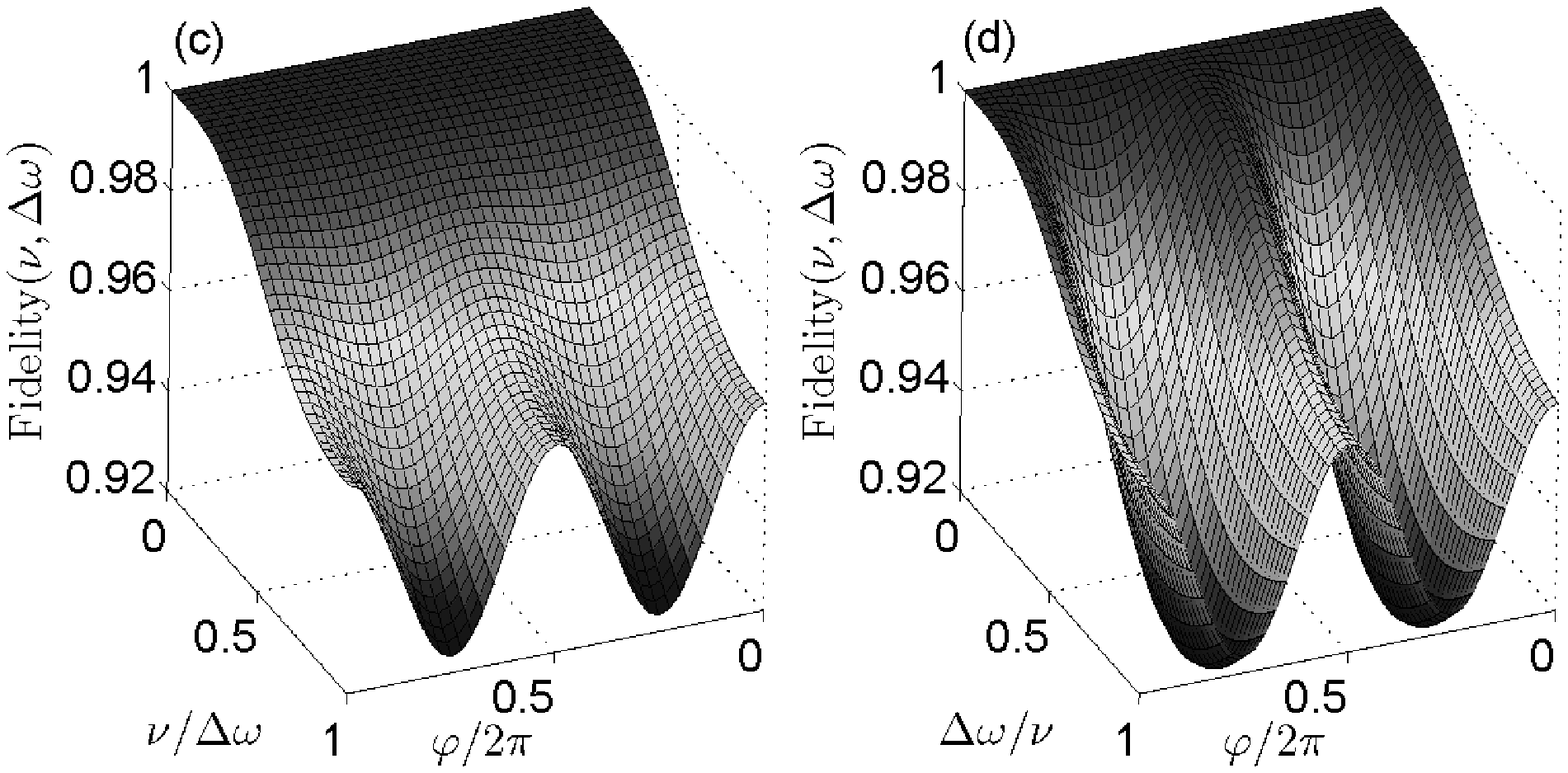}
\caption{{\protect\footnotesize The fidelity of single-qubit gate in
 the presence of
fluctuations on both $\Delta\protect\omega$ and $\protect\nu$, where $\protect%
\varphi_{i}=0$ in (a) and (b), $\protect\theta_{i}=\protect\pi/2$ in
(c) and (d). Parameters are: $\omega_{10}/2\pi=6$GHz,
$\protect\sigma%
_{0}=0.1$, and $\protect\sigma_{1}=0.1$.}}
\label{fig:nu_delta_omega}
\end{figure}
\subsection{Fidelity of single-qubit gates}

In the numerical simulations reported here, we randomly choose more
than ten thousand stochastic numbers ($N\geq10000$) for a given mean
squared noise $\sigma$
(for brevity but without loss of generality, we hereafter set $\tilde{\eta}%
=\eta=0$ and neglect its randomness). We select the experimental
parameter of superconducting phase qubits: $\omega_{10}/2\pi=6$GHz
and $\sigma_{0}=\sigma_{1}=0.1$, where $\sigma_{0}$ and $\sigma_{1}$
represents the fluctuation of $\Delta\omega$ and $\nu$ respectively.
We then calculate the average fidelity (up to satisfactory
convergence) versus the coordinates of the initial state and the
parameters, as depicted
in figures~\ref{fig:delta_omega}, \ref%
{fig:nu} and \ref{fig:nu_delta_omega}, respectively. Several
remarkable features can be seen from the figures. (i) The calculated
fidelity of geometric quantum gates for any input state
 is rather high (larger than 0.92) for the considered noises. Actually, the amplitude
of microwave current could be controlled precisely in experiments.
Therefore, the $\sigma_{1}$ is much smaller than $\sigma_{0}$ in
superconducting phase qubits. From the above disscusions, if the
$\sigma_{0}$ is about 0.03,  the designed geometric quantum gate
is likely rather insensitive to the stochastic errors.
 (ii)The suppression effect of $
\Delta\omega$ fluctuations on the fidelity is weaker than that of
$\nu$ fluctuations, which becomes more pronounced when the mean
squared noise
is stronger(not shown here). Also reasonablly, the joint effect of both $\Delta\omega$ and $%
\nu$ fluctuations on the fidelity is relatively stronger than any
single one, but the shapes of the cooresponding figures are
similar. Note that, since it seems difficult to control the
$\Delta\omega$ precisely because $\Delta\omega$ often varies due
to the noise current in experiments~\cite{Martinis2003},
 to optimize a quantum gate with the persent
geometric scenario may be quite helpful. (iii) The fidelity is very
close to 1 for $\nu \ll\Delta\omega$ or $\nu\gg\Delta\omega$.
Actually, we have a trivial geometric phase $2\pi$ and a trivial
unit gate in this case.
(iv) For a given $\xi$, the fidelity reaches a maximum when the
input state is the eigenstate of the Hamiltonian.

\subsection{Fidelity of two-qubit
gates} We now turn to address a kind of non-trivial two-qubit
controlled phase gate in the present system. So far, many efforts
have been paid to a  two-qubit controlled phase gate with
\textbf{$\sigma_{z}\sigma_{z}$} coupling\cite{Falci,Peng,Jones,ZW}.
However, the present two superconducting phase qubits are coupled
with a capacitor [see figure~\ref{fig:circuit}(b)], and thus the
qubit-qubit interaction takes the \textbf{$\sigma_{y}\sigma_{y}$}
form, with the total Hamiltonian being given by
\begin{equation}
\hat{H}={\textstyle \sum_{i=a,b}} \hat{H}_{i}+{\frac{J}{2} {\hat{\sigma}_{y}}%
^{(a)} {\hat{\sigma}_{y}}^{(b)},}
\end{equation}
where the coupling strength $J\approx(C_{x}/C_{J})\hbar\omega_{01}$.
This Hamiltonian could be used to manipulate the target qubit
(\emph{qubit b}) for the realization of a two-qubit gate under the
condition that the control qubit (\emph{qubit a}) is off resonance
in the operation of the target qubit. The Hamiltonian of \emph{qubit
b} in the {$\omega_{10}^b$} rotating frame is dependent on the state
of \emph{qubit b} through the coupling term $J$:  the contribution
is $J/2$ (or $-J/2$) if the state of \emph{qubit a} is
$|\psi_a\rangle=|-\rangle$ (or $|\psi_a\rangle=|+\rangle$), with
$|-\rangle$ and $|+\rangle$ as the two eigenstates of $\sigma_y$.
Setting  $\tilde{\eta}=\eta=\pi/2$ and after an adiabatic evolution
loop, the acquired geometric phase of target qubit is derived as
\begin{eqnarray}\label{twoqubit}
\gamma_{g}^+=\frac{1}{2}\int_{\frac{\pi}{2}}^{\frac{5\pi}{2}}\frac{\nu^2-\frac{1}{2}a\sin\phi}{(\sqrt{-a\sin\phi+b})(\sqrt{-a\sin\phi+b}+\Delta\omega)}d\phi,\nonumber\\
\gamma_{g}^-=\frac{1}{2}\int_{\frac{\pi}{2}}^{\frac{5\pi}{2}}\frac{\nu^2+\frac{1}{2}a\sin\phi}{(\sqrt{a\sin\phi+b})(\sqrt{a\sin\phi+b}+\Delta\omega)}d\phi,
\end{eqnarray}
where $a=\nu J, b=\nu^2+\Delta\omega^2+\frac{1}{4}J^2$. Although
$\gamma_{g}^+=\gamma_{g}^-=\gamma/2$, we can still have a nontrivial
two-qubit controlled geometric phase gate, given by
\begin{equation}
\label{equation:control} U_{ctrl}=\left(
\begin{array}{cc}
U_{(\gamma, \xi^{+})} &
0  \\
0 & U_{(\gamma, \xi^{-})}%
\end{array}%
\right),
\end{equation}
where $\xi^{\pm}=\tan^{-1}[(\nu\mp J/2)/\Delta\omega]$.

\begin{figure}
\centering
\includegraphics[width=0.5\textwidth]{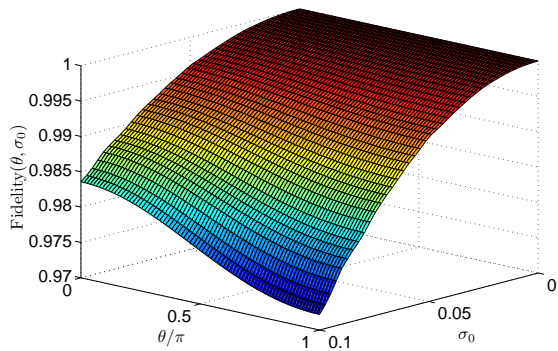}
\caption{{\protect\footnotesize } The fidelity of the two-qubit gate
in the presence of the $\Delta\omega$ fluctuation with various
$\sigma_{0}$ when the input state is
$(\cos\frac{\theta_{i}}{2}|+\rangle+\sin\frac{\theta_{i}}{2}|-\rangle)_{C}\otimes|0\rangle_{T}$.
Parameters are: $\omega_{10}^a/2\pi\neq\omega_{10}^b/2\pi=6$GHz,
$\Delta\omega/2\pi=\nu/2\pi=300$MHz, and $J/2\pi=150$MHz. }
\label{fig:twoqubit}
\end{figure}

We now numerically simulate the fidelity of the two-qubit gate by
assuming only $\Delta\omega$ fluctuations, which are believed to be
more significant than $\nu$ fluctuations in superconducting phase
qubits. The input state is
$(\cos\frac{\theta_{i}}{2}|+\rangle+\sin\frac{\theta_{i}}{2}|-\rangle)_{C}\otimes|0\rangle_{T}$,
with subscripts C and T mean the state of the control and target
qubits, respectively.
Indeed, for the typical experiment parameters:
$\omega_{10}/2\pi\sim6$GHz,
$\Delta\omega/2\pi=\nu/2\pi=300$MHz,$C_x\sim33$fF, $C_J\sim1.3$pF in
Ref.\cite{Steffen,Martinis2005,Martinis2006}, we have
$J/2\pi\sim150$MHz. Therefore, the fidelity of the two-qubit gate in
the presence of $\Delta\omega$ fluctuation with various $\sigma_{0}$
is shown in figure~\ref{fig:twoqubit}. The fidelities of the
two-qubit controlled phase gate are larger than 0.972. It seems that
the fidelity of two-qubit gate is also robust against the random
errors resulting from thermal fluctuation in superconducting phase
qubits as well.

\section{Detection of Berry Phase in Superconducting Phase Qubits}
\label{section:detection} Supercondcuting qubits, considered as
artificial macroscopic two-level atoms, are also proposed as a
candidate for detecting the geometric phase in macroscopic quantum
systems\cite{Falci,Peng}. However, these existing  proposals suggest
to detect the Berry phase through the interference measurement, in
which the dephasing may affect seriously the visibility in measuring
this phase. In particular, it seems quite difficult to detect the
Berry phase via the interference measurement in superconducting
phase qubits.
\begin{figure}
\centering
\includegraphics[width=0.45\textwidth]{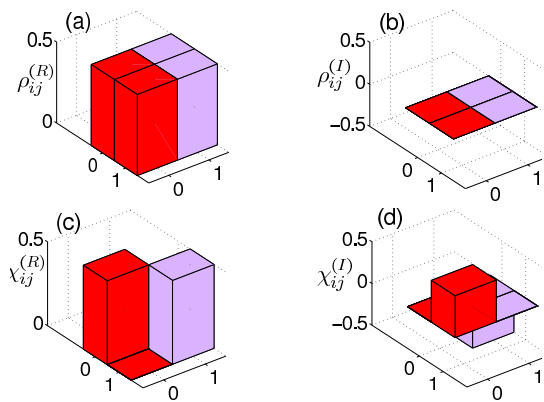}
\caption{{\protect\footnotesize Graphical representation of the
density matrices $\protect\rho$ and $\protect\chi$ for the initial
and final states,  with (a)\&(c) and (b)\&(d) denoting respectively
the real and imaginary parts, where $\protect\theta=\protect\pi/2$ and $\protect\gamma_{g}^{0}=\protect%
\pi/8$. }} \label{fig:single}
\end{figure}
 Recently,
Steffen \textit{et al.}\cite{Steffen} reported the first
demonstration of quantum state tomography using single shot
measurements in superconducting phase qubits. Stimulated by this
experiment, we here propose to  directly detect the adiabatic Berry
phase and to measure the fidelity of the designed geometric quantum
gate via the quantum
state tomography in future experiments. 

Let us illustrate an example below. The initial state of qubit is
prepared
as $|\psi_{i}\rangle=[\cos(\theta/2),
\sin(\theta/2)]^{T}$ in the basis of [$|\psi_{+}\rangle$, $|\psi_{-}\rangle$%
] (i.e., [$|0\rangle$, $|1\rangle$] if we set $\nu(0)=0$). As we
described before, we drive the field to loop twice in a designated
way in
the parameter space, 
and thus the final state is $|\psi_{f}\rangle=[e^{2i\gamma^0_g}\cos(%
\theta/2),e^{ -2i\gamma^0_g}\sin(\theta/2)]^{T}$. 
One rotates the Bloch vector of the qubit state in Bloch sphere with
microwave current pulses along x, y and z directions and measures
the first excitation $|\psi_{-}\rangle\langle \psi_{-}|$ for
reconstructing the density matrix of the state. The relative phase
change between $|\psi_{+}\rangle$ and $|\psi_{-}\rangle$ observed
through the state tomography is $4\gamma^0_g$. The corresponding
qubit state can be graphically represented, as shown in
figure~\ref{fig:single}. From this figure, the Berry phase may be
determined from the relative phase of the density matrix elements in
(or between) the final state (and the initial state). In
experiments, one may optimize the experimental result by some
methods which help to reduce the unavoidable decoherence and
statistical errors \cite{Paris}. In addition, the target qubit
conditional phase shift may be detected by the simultaneous joint
measurement of two-qubit state \cite{Martinis2005,Martinis2006}.

\section{Discussions and Summary}
\label{section:discussion} The experimental realization of our
proposal for Adiabatic Abelian gates and detecting the Berry phase
in superconducting phase qubits is quite possible, although it may
meet various technological challenges.  The rapid inversion of the
bias magnetic field to cancel the the dynamical contribution to the
overall phase is experimentally feaisble. In fact, with a
flux-biased superconducting phase qubit (which is essentially a
current-biased Josephson junction) loop size of 50 $(\mu
m)^2$\cite{Martinis2006,Martinis2004}, changing the flux by about
half of a flux quantum in $10^{-10}$s, requires sweeping the
magnetic field at a rate of about $2\times10^5$ T/s, that is
reachable by current techniques\cite{Uwazumi}.

Perhaps a main challenge is the implementation of the adiabatic
evolution of the Hamiltonian to get the Berry phase within the
qubit¡¯s decoherence time, which in turn must be longer than the
typical timescale of superconducting phase qubits: $2\pi/\omega
_{10}$, $2\pi/(\omega_{10}-\omega_{21})\thicksim$3 ns. The slowly
varying the phase of microwave current could be realized with 100
linear steps of about 4 ns. The required microwave technique is
rather mature \cite{Steffen,Martinis2006}. In view of the
decoherence time data in Refs.\cite{Yu} and \cite{Steffen}, it
seems feasible to detect the geometric phase in phase qubits with
the current quantum state tomography technology.

As for a direct comparison of the Adiabatic Abelian gates and the
dynamic gates in superconducting qubits, to our knowledge, it was
indicated before that  the geometric gates are more robust against
fluctuations of control parameters than dynamic gates
~\cite{Zhu2006}. From our numerical simulations, adiabatic Abelian
geometric gates are likely robust against certain random errors in
superconducting phase qubits.  We wish to indicate that although
there are limitations caused by the adiabatic condition, geometric
QC based on the adiabatic Berry phase may have an interesting
application in a precise preparation of a quantum
state\cite{Peng,Zhu2006}, mainly due to its global geometric
robustness against certain kind of errors. An experimental process
to determine the noisy channel of the controlled qubits based on
the qubit state tomography is referred to as quantum process
tomography \cite{Nielsen,Liu}. Since a set of standard qubit
states are required to be precisely prepared in the quantum
process tomography experiments, we may use the present geometric
QC strategy to achieve them. For example, to realize quantum
process
tomography for a single phase qubit, the four kinds of input states $%
|0\rangle,|1\rangle, |+\rangle=(|0\rangle +|1\rangle)/\sqrt2$ and $%
|-\rangle=(|0\rangle+i|1\rangle)/\sqrt2$ need to be precisely
prepared. In our scheme, the state $|-\rangle$ can be made from an
easy initial state $|0\rangle$ once we set
 $%
\varphi_{i}=0$, $\theta_{i}=0$, $\xi=\pi/2$, and
$\gamma_g^{0}=\pi/8$, with a relative high fidelity for weaker
noises.

In summary, we have developed an adiabatic Abelian geometric QC
strategy based on the non-degenerate energy eigenstates. The
fidelity of the designed quantum gate has been evaluated in the
presence of simulated Gaussian-type thermal fluctuation noises in
superconducting phase qubits and found to be rather robust against
the random errors. A possible application of our strategic scheme in
a precise preparation of designated quantum state has been
addressed. We have also proposed to detect directly the Berry phase
in phase qubits via the quantum state tomography.

\section*{Acknowledgments}
\addcontentsline{toc}{section}{Acknowledgments}

 We thank S.Y. Han, B. Xiong for
useful discussions and B.Y. Zhu for kind help. Peng thanks P. Leek
and A. Wallraff for showing their latest results prior to
publication. This work was supported by the National Natural Science
Foundation of China (10534060, 10574154, 10221002, and 10429401),
the Ministry of Science and Technology of China through the 973 and
the state key programs (2006CB601007, 2006CB921107, 2006CB0L1001),
the Chinese Academy of Sciences, the RGC of Hong Kong (HKU 7045/05P,
HKU-3/05C, HKU 7049/07P), and the URC fund at University of Hong
Kong.

\emph{Note added.}---After completion of this work, we learned that
Wallraff \emph{et al.} have observed the Berry phase superconducting
charge qubits with microwave techniques via quantum state tomography
which are similar with the idea proposed in the paper
\cite{Wallraff2007}.

\end{document}